\documentclass[conference]{IEEEtran}
\IEEEoverridecommandlockouts
\usepackage{cite}
\usepackage{amsmath,amssymb,amsfonts}
\usepackage{algorithmic}
\usepackage{graphicx}
\usepackage{textcomp}
\usepackage{xcolor}
\usepackage{multirow}
\usepackage[super]{nth}
\def\BibTeX{{\rm B\kern-.05em{\sc i\kern-.025em b}\kern-.08em
    T\kern-.1667em\lower.7ex\hbox{E}\kern-.125emX}}
\begin{document}

\title{Channel Optimized Visual Imagery based \\ Robotic Arm Control under the Online Environment
\footnote{
{\thanks{20xx IEEE. Personal use of this material is permitted. Permission
from IEEE must be obtained for all other uses, in any current or future media, including reprinting/republishing this material for advertising or promotional purposes, creating new collective works, for resale or redistribution to servers or lists, or reuse of any copyrighted component of this work in other works.}}
{\thanks{This work was partly supported by Institute of Information \& Communications Technology Planning \& Evaluation (IITP) grant funded by the Korea government (MSIT) (No. 2017-0-00432, Development of Non-Invasive Integrated BCI SW Platform to Control Home Appliances and External Devices by User’s Thought via AR/VR Interface; No. 2017-0-00451, Development of BCI based Brain and Cognitive Computing Technology for Recognizing User’s Intentions using Deep Learning; No. 2019-0-00079, Artificial Intelligence Graduate School Program, Korea University).}
}}
}

\author{\IEEEauthorblockN{Byoung-Hee Kwon}
\IEEEauthorblockA{\textit{Dept. Brain and Cognitive Engineering} \\
\textit{Korea University}\\
Seoul, Korea \\
bh\_kwon@korea.ac.kr}

\and

\IEEEauthorblockN{Byeong-Hoo Lee}
\IEEEauthorblockA{\textit{Dept. Brain and Cognitive Engineering} \\
\textit{Korea University}\\
Seoul, Korea \\
bh\_lee@korea.ac.kr}

\and

\IEEEauthorblockN{Jeong-Hyun Cho}
\IEEEauthorblockA{\textit{Dept. Brain and Cognitive Engineering} \\
\textit{Korea University}\\
Seoul, Korea \\
jh\_cho@korea.ac.kr}

}

\maketitle

\begin{abstract}
An electroencephalogram is an effective approach that provides a bidirectional pathway between the user and computer in a non-invasive way. In this study, we adopted the visual imagery data for controlling the BCI-based robotic arm. Visual imagery increases the power of the alpha frequency range of the visual cortex over time as the user performs the task. We proposed a deep learning architecture to decode the visual imagery data using only two channels and also we investigated the combination of two EEG channels that has significant classification performance. When using the proposed method, the highest classification performance using two channels in the offline experiment was 0.661. Also, the highest success rate in the online experiment using two channels (AF3--Oz) was 0.78. Our results provide the possibility of controlling the BCI-based robotic arm using visual imagery data.
\\
\end{abstract}

\begin{small}
\end{small}

\section{INTRODUCTION}
The brain-computer interface (BCI) allows users to communicate with computers using brain signals \cite{vaughan2003brain, zhang2019strength, kwon2019subject, lee2019connectivity}. Electroencephalography (EEG) has the advantage of having a higher time resolution than comparable methods like near-infrared spectroscopy \cite{chen2016high, lee2017network} and functional magnetic resonance imaging (fMRI) \cite{zhang2017hybrid}. This study applied an endogenous paradigm based on visual imagery for EEG-based BCI.

Various studies have been conducted to decode human intentions based on brain signals or other bio-signals in the last few years \cite{jeong2018decoding, he2018brain, lee2020continuous, chholak2019visual, suk2014predicting, thung2018conversion, kim2019subject}. In order to control BCI-related devices, EEG signals associated with the user's intentions were analyzed using several BCI paradigms. P300 \cite{yeom2014efficient, lee2018high, won2017motion}, steady-state visual evoked potentials (SSVEP) \cite{won2015effect, kwak2017convolutional}, and motor imagery (MI) \cite{kim2014decoding, zhang2021adaptive, kam2013non, jeong2020decoding} were implemented to control BCI-related devices. Using exogenous paradigms such as SSVEP and P300 can decrease concentration and fatigue among users because they require external devices. Furthermore, MI is perceived differently by each individual, which results in a lack of consistency. This may result in discrepancies between the user's intentions and the actual outcome.

We used visual imagery in this study to overcome these limitations. The user performs visual imagery when they visualize a picture or movement as if they were drawing a picture. Visual imagery is a paradigm based on visual perception experiences without the need for additional external devices \cite{kwon2020decoding}. As a result of visual imagery, a wide range of brain signals are generated from the frontal and occipital areas, containing the visual cortex. It is possible to analyze visual imagery in a variety of frequency ranges, including delta, theta, and alpha bands, and the prefrontal and occipital lobes are mainly activated \cite{koizumi2018development}. It is clear that visual imagery visual perception can be decoded in the visual cortex, including V1 and V2, through activities based on visual imagery \cite{sousa2017pure}. These activities induce delta bands in the prefrontal lobes and alpha bands in the occipital lobes.

\begin{figure}[t!]

\includegraphics[scale = 0.9]{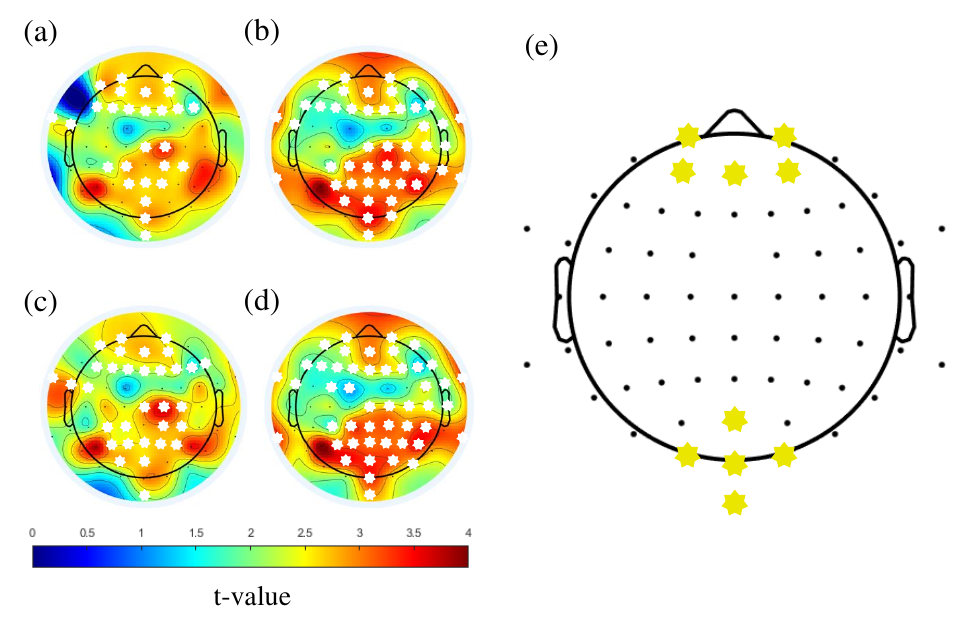}
\caption{Permutation test results. The intensity of activation was expressed as \textit{t}-values. White asterisks indicate electrodes that are significantly different between the imagery phase and the rest phase (\textit{p} $\leq$ 0.01). In the following list of examples, (a) through (d) means pouring water, opening the door, eating food, and picking up a cell phone, respectively. (e) shows the most significant channels based on statistical analysis and previous studies.}
\label{fig:1}
\end{figure}

Upon looking at an object, a specific brain signal is manifested in the visual cortex, which is known as visual perception. During visual imagery, brain signals follow a similar path to visual perception, and their intensity increases as time passes. In the visual cortex, visual perception leads to a reduction in brain activity within the alpha frequency range over time, and as the user continues to do the task, visual imagery leads to an increase in the alpha frequency range in the visual cortex. This study aimed to reduce the differences between visual perception and visual imagery and construct a neural network accordingly to improve visual imagery decoding. As part of the visual imagery classes, the user was asked to complete four tasks (pouring water, opening a door, eating food, and picking up a cell phone), while performing visual imagery, they were instructed to perform a task on the black screen to demonstrate the difference between visual perception and visual imagery. In this study, we were able to identify the most meaningful channels for visual imagery and confirmed the possibility of controlling a BCI-based robotic arm in real-life.

\section{METHODS}
\subsection{Dataset}
The EEG data of eight subjects from our previous study (S01-S08); ages 24-30 (Mean: 26.6, SD: 1.89; 4 men and 4 women, all right-handed) were used. We used 64 Ag/AgCl electrodes with a 1,000 Hz sampling rate (Fp1--2, AF3--4, AF7--8, AFz, F1--8, Fz, FC1--6, FT7--10, C1--6, Cz, T7--8, CP1--6, CPz, TP7--10, P1-8, PZ, PO3-4, PO7--8, O1--2, Oz, Iz) via BrainAmp (BrainProduct GmbH, Germany) via a 10/20 international system (BrainProduct GmbH, Germany). 

For the purpose of acquiring good-quality visual imagery-related EEG signals, visual imagery paradigm consists of three stages: the rest stage, the instruction stage, and the visual imagery stage. After the visual imagery stage, there is a 5-s pause between the visual stimulus and the rest stage so that the aftereffect of previous visual stimuli are not experienced. The data were collected from 200 trials per subject, of which 50 trials were collected for each class. A visual imagery task consists of pouring water, opening the door, eating food, and picking up a cell phone.


\begin{figure}[t!]
\centering
\includegraphics[width=\columnwidth]{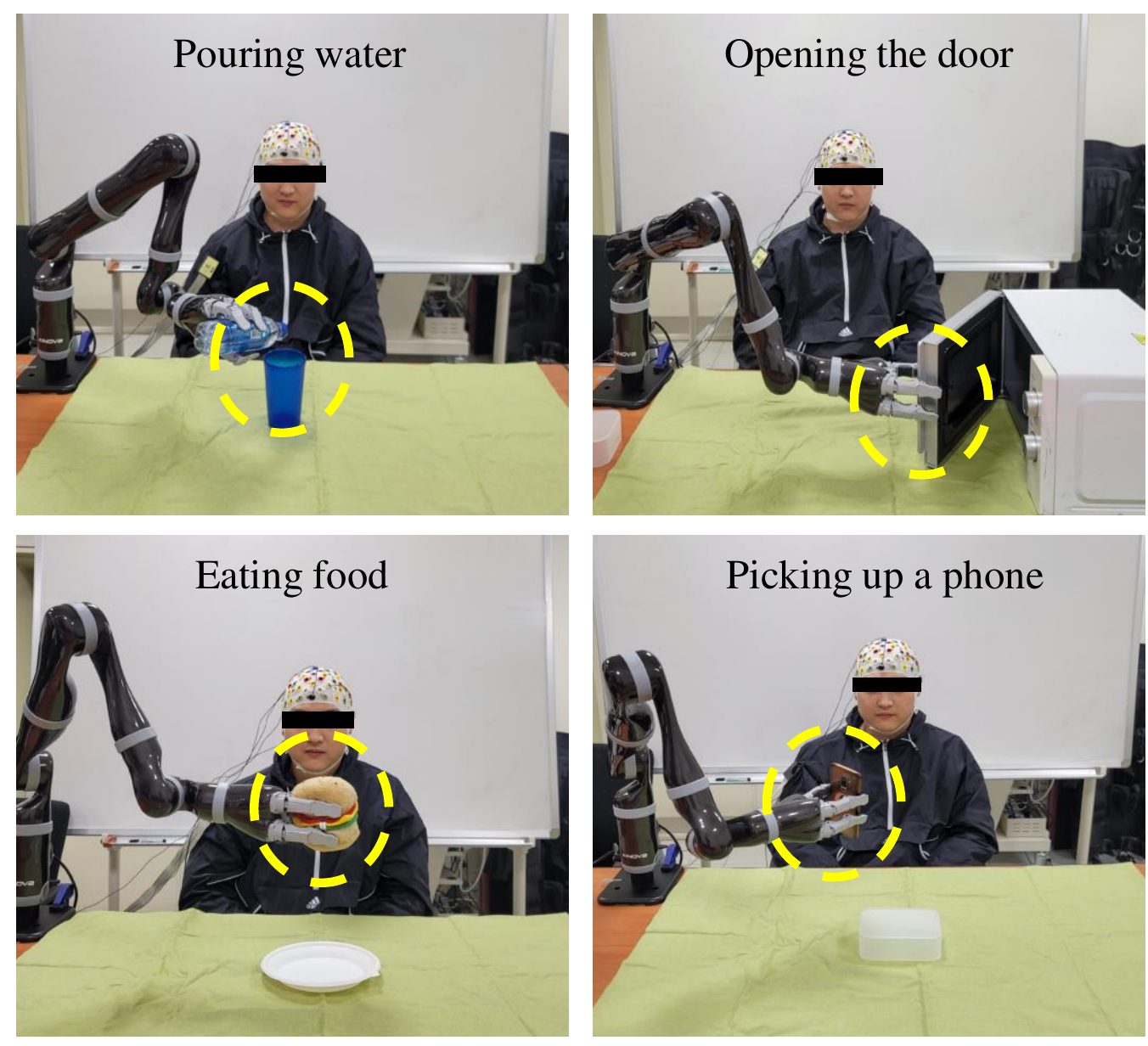}
\caption{The environment of online BCI-based robotic arm control system. Each class consists of 10 trials, and the user performed a total of 40 visual imagery trials. Yellow circles indicate the class that the robot arm performed.}
\end{figure}

\subsection{Data Analysis}
 To preprocess the data, BBCI toolbox and openBMI \cite{lee2019eeg} were used with MATLAB 2020a (MathWorks Inc., USA). In visual imagery, the band-pass filter was applied between [0.5--13] Hz, corresponding to significant frequencies such as delta, theta, and alpha. Based on a one-versus-rest approach, we selected the significant channels for controlling the BCI-based robotic arm in the online environment. Also, we investigated important channels in the visual imagery task through spatial comparison with significant differences in brain activation, to consider the practicality of BCI-related devices.

\begin{table*}[t!]
\caption{Performances of Visual Imagery Classification with Significant Channels}
\renewcommand{\arraystretch}{1.2}

\begin{tabular*}{\textwidth}{@{\extracolsep{\fill}\quad}ccccccccccc}
\hline
      & Fp1                                                      & Fp2                                                      & AFz                                                      & AF3                                                      & AF4                                                      & POz                                                       & O1                                                       & O2                                                       & Oz                                                       & Iz                                                       \\ \hline
Sub01 & \begin{tabular}[c]{@{}c@{}}0.591\\ (±0.014)\end{tabular} & \begin{tabular}[c]{@{}c@{}}0.603\\ (±0.019)\end{tabular} & \begin{tabular}[c]{@{}c@{}}0.632\\ (±0.011)\end{tabular} & \begin{tabular}[c]{@{}c@{}}0.611\\ (±0.026)\end{tabular} & \begin{tabular}[c]{@{}c@{}}0.597\\ (±0.012)\end{tabular} & \begin{tabular}[c]{@{}c@{}}0.581\\ (±0.010)\end{tabular} & \begin{tabular}[c]{@{}c@{}}0.609\\ (±0.018)\end{tabular} & \begin{tabular}[c]{@{}c@{}}0.650\\ (±0.023)\end{tabular} & \begin{tabular}[c]{@{}c@{}}0.577\\ (±0.008)\end{tabular} & \begin{tabular}[c]{@{}c@{}}0.589\\ (±0.016)\end{tabular} \\
Sub02 & \begin{tabular}[c]{@{}c@{}}0.688\\ (±0.011)\end{tabular} & \begin{tabular}[c]{@{}c@{}}0.695\\ (±0.016)\end{tabular} & \begin{tabular}[c]{@{}c@{}}0.689\\ (±0.029)\end{tabular} & \begin{tabular}[c]{@{}c@{}}0.706\\ (±0.012)\end{tabular} & \begin{tabular}[c]{@{}c@{}}0.662\\ (±0.006)\end{tabular} & \begin{tabular}[c]{@{}c@{}}0.683\\ (±0.005)\end{tabular} & \begin{tabular}[c]{@{}c@{}}0.61\\ (±0.003)\end{tabular}  & \begin{tabular}[c]{@{}c@{}}0.615\\ (±0.027)\end{tabular} & \begin{tabular}[c]{@{}c@{}}0.709\\ (±0.030)\end{tabular} & \begin{tabular}[c]{@{}c@{}}0.631\\ (±0.020)\end{tabular} \\
Sub03 & \begin{tabular}[c]{@{}c@{}}0.596\\ (±0.010)\end{tabular} & \begin{tabular}[c]{@{}c@{}}0.628\\ (±0.021)\end{tabular} & \begin{tabular}[c]{@{}c@{}}0.617\\ (±0.010)\end{tabular} & \begin{tabular}[c]{@{}c@{}}0.629\\ (±0.011)\end{tabular} & \begin{tabular}[c]{@{}c@{}}0.580\\ (±0.013)\end{tabular} & \begin{tabular}[c]{@{}c@{}}0.576\\ (±0.020)\end{tabular} & \begin{tabular}[c]{@{}c@{}}0.612\\ (±0.012)\end{tabular} & \begin{tabular}[c]{@{}c@{}}0.579\\ (±0.029)\end{tabular} & \begin{tabular}[c]{@{}c@{}}0.643\\ (±0.022)\end{tabular} & \begin{tabular}[c]{@{}c@{}}0.625\\ (±0.005)\end{tabular} \\
Sub04 & \begin{tabular}[c]{@{}c@{}}0.563\\ (±0.020)\end{tabular} & \begin{tabular}[c]{@{}c@{}}0.538\\ (±0.012)\end{tabular} & \begin{tabular}[c]{@{}c@{}}0.569\\ (±0.007)\end{tabular} & \begin{tabular}[c]{@{}c@{}}0.589\\ (±0.027)\end{tabular} & \begin{tabular}[c]{@{}c@{}}0.536\\ (±0.008)\end{tabular} & \begin{tabular}[c]{@{}c@{}}0.541\\ (±0.011)\end{tabular} & \begin{tabular}[c]{@{}c@{}}0.577\\ (±0.026)\end{tabular} & \begin{tabular}[c]{@{}c@{}}0.550\\ (±0.020)\end{tabular} & \begin{tabular}[c]{@{}c@{}}0.514\\ (±0.018)\end{tabular} & \begin{tabular}[c]{@{}c@{}}0.512\\ (±0.002)\end{tabular} \\
Sub05 & \begin{tabular}[c]{@{}c@{}}0.618\\ (±0.017)\end{tabular} & \begin{tabular}[c]{@{}c@{}}0.571\\ (±0.021)\end{tabular} & \begin{tabular}[c]{@{}c@{}}0.572\\ (±0.009)\end{tabular} & \begin{tabular}[c]{@{}c@{}}0.594\\ (0.027±)\end{tabular} & \begin{tabular}[c]{@{}c@{}}0.602\\ (±0.008)\end{tabular} & \begin{tabular}[c]{@{}c@{}}0.616\\ (±0.026)\end{tabular} & \begin{tabular}[c]{@{}c@{}}0.606\\ (±0.011)\end{tabular} & \begin{tabular}[c]{@{}c@{}}0.581\\ (±0.025)\end{tabular} & \begin{tabular}[c]{@{}c@{}}0.621\\ (±0.020)\end{tabular} & \begin{tabular}[c]{@{}c@{}}0.594\\ (±0.010)\end{tabular} \\
Sub06 & \begin{tabular}[c]{@{}c@{}}0.582\\ (±0.007)\end{tabular} & \begin{tabular}[c]{@{}c@{}}0.615\\ (±0.029)\end{tabular} & \begin{tabular}[c]{@{}c@{}}0.615\\ (±0.028)\end{tabular} & \begin{tabular}[c]{@{}c@{}}0.607\\ (±0.015)\end{tabular} & \begin{tabular}[c]{@{}c@{}}0.611\\ (±0.013)\end{tabular} & \begin{tabular}[c]{@{}c@{}}0.611\\ (±0.005)\end{tabular} & \begin{tabular}[c]{@{}c@{}}0.585\\ (±0.021)\end{tabular} & \begin{tabular}[c]{@{}c@{}}0.603\\ (±0.008)\end{tabular} & \begin{tabular}[c]{@{}c@{}}0.601\\ (±0.020)\end{tabular} & \begin{tabular}[c]{@{}c@{}}0.612\\ (±0.015)\end{tabular} \\
Sub07 & \begin{tabular}[c]{@{}c@{}}0.585\\ (±0.013)\end{tabular} & \begin{tabular}[c]{@{}c@{}}0.577\\ (±0.014)\end{tabular} & \begin{tabular}[c]{@{}c@{}}0.604\\ (±0.018)\end{tabular} & \begin{tabular}[c]{@{}c@{}}0.573\\ (±0.010)\end{tabular} & \begin{tabular}[c]{@{}c@{}}0.606\\ (±0.008)\end{tabular} & \begin{tabular}[c]{@{}c@{}}0.583\\ (±0.019)\end{tabular} & \begin{tabular}[c]{@{}c@{}}0.590\\ (±0.017)\end{tabular} & \begin{tabular}[c]{@{}c@{}}0.601\\ (±0.030)\end{tabular} & \begin{tabular}[c]{@{}c@{}}0.572\\ (±0.015)\end{tabular} & \begin{tabular}[c]{@{}c@{}}0.594\\ (±0.011)\end{tabular} \\
Sub08 & \begin{tabular}[c]{@{}c@{}}0.563\\ (±0.018)\end{tabular} & \begin{tabular}[c]{@{}c@{}}0.540\\ (±0.019)\end{tabular} & \begin{tabular}[c]{@{}c@{}}0.577\\ (±0.027)\end{tabular} & \begin{tabular}[c]{@{}c@{}}0.579\\ (±0.018)\end{tabular} & \begin{tabular}[c]{@{}c@{}}0.54\\ (±0.017)\end{tabular}  & \begin{tabular}[c]{@{}c@{}}0.532\\ (±0.030)\end{tabular} & \begin{tabular}[c]{@{}c@{}}0.531\\ (±0.015)\end{tabular} & \begin{tabular}[c]{@{}c@{}}0.567\\ (±0.013)\end{tabular} & \begin{tabular}[c]{@{}c@{}}0.579\\ (±0.011)\end{tabular} & \begin{tabular}[c]{@{}c@{}}0.572\\ (±0.023)\end{tabular} \\ \hline
Avg.  & 0.598                                                    & 0.595                                                    & 0.609                                                    & 0.611                                                    & 0.591                                                    & 0.590                                                    & 0.590                                                    & 0.593                                                    & 0.602                                                    & 0.591                                                    \\ \hline
\end{tabular*}
\end{table*}

\subsection{Channel optimization method}
In this paper, we investigated optimized EEG channels when the subjects performed the visual imagery tasks to control the BCI-based robotic arm in an online environment. Using deep learning approaches based on convolutional neural networks (CNN) that consists of 3 convolution layers, we identified channels that were appropriate for online application based on their results. With an optimized order (\textit{N}= 30), Hamming-windowed zero phase finite impulse response (FIR) filters were used to band-pass filter the EEG data between 0.5 and 13 Hz, focusing on the delta, theta, and alpha frequencies that are associated with visual imagery. We used a sliding window as an augmentation method with a length of 2 seconds and a 50 \% overlap to increase the amount of training data for a deep learning network. To begin training the networks, 80 \% of the trials were randomly chosen for training, and 20 \% were selected for performance evaluation. The training is done over 200 epochs, with 16 batches, and a learning rate of 0.0001.

\subsection{Online experiment}
An online experiment was conducted to verify that the BCI-based robotic arm was feasible. The user sat in a comfortable position about 30cm away from the robotic arm and performed the visual imagery task with a JACO arm (KINOVA Inc., Canada). Our method of obtaining EEG signals with only two channels was optimized based on the results obtained from the channel optimization method we performed previously. Using a CNN-based deep learning network in offline experiments, all settings were set the same for analyzing user intentions.

\section{RESULTS and DISCUSSION}

\begin{table}[t!]
\tiny
\caption{Performances of Visual Imagery Classification with Combinations of Significant Channels}
\renewcommand{\arraystretch}{1.3}
\begin{center}
\resizebox{\columnwidth}{!}{%
\begin{tabular}{ccccc}
\hline
      & AFz--AF3 & AFz--Oz & AF3--Oz & Avg.  \\ \hline
Sub01 & 0.687   & 0.642  & 0.627  & 0.652 \\
Sub02 & 0.638   & 0.633  & 0.656  & 0.642 \\
Sub03 & 0.661   & 0.641  & 0.639  & 0.647 \\
Sub04 & 0.592   & 0.609  & 0.631  & 0.610 \\
Sub05 & 0.682   & 0.616  & 0.658  & 0.652 \\
Sub06 & 0.636   & 0.698  & 0.654  & 0.663 \\
Sub07 & 0.670   & 0.675  & 0.656  & 0.667 \\
Sub08 & 0.673   & 0.682  & 0.641  & 0.665 \\ \hline
\end{tabular}
}
\end{center}
\end{table}
\subsection{Data analysis}
The significance of brain activation differences between each class and the rest class was examined based on one versus rest approaches. Using statistical analysis, Fig. 1 depicts the spatial differences in spectral power between each class and resting state. The prefrontal and occipital lobes showed significant activity in the study, while other brain regions did not show statistically significant differences. Using these results and previous studies that indicated significant EEG channels in visual imagery, we selected 10 EEG channels (Fp1--2, AFz, AF3--4, POz, Oz, O1--2, Iz) to decode the visual imagery data.

\begin{table}[t!]
\tiny
\caption{Evaluation Performance for Online Experiment Analysis \\ through the Success Rate of Decoding}
\renewcommand{\arraystretch}{1.3}
\begin{center}
\resizebox{\columnwidth}{!}{%
\begin{tabular}{ccccc}
\hline
                       &      & AFz-AF3      & AF3-Oz       & AFz-Oz       \\ \hline
\multirow{3}{*}{Sub06} & Run1 & 0.73 (29/40) & 0.68 (27/40) & 0.68 (27/40) \\
                       & Run2 & 0.60 (24/40) & 0.63 (25/40) & 0.70 (28/40) \\
                       & Run3 & 0.70 (28/40) & 0.65 (26/40) & 0.55 (22/40) \\ \hline
\multirow{3}{*}{Sub07} & Run1 & 0.55 (22/40) & 0.63 (25/40) & 0.73 (29/40) \\
                       & Run2 & 0.70 (28/40) & 0.78 (31/40) & 0.68 (27/40) \\
                       & Run3 & 0.63 (25/40) & 0.60 (24/40) & 0.60 (24/40) \\ \hline
\multirow{3}{*}{Sub08} & Run1 & 0.70 (28/40) & 0.63 (25/40) & 0.55 (22/40) \\
                       & Run2 & 0.75 (30/40) & 0.68 (27/40) & 0.58 (23/40) \\
                       & Run3 & 0.60 (24/40) & 0.75 (30/40) & 0.55 (22/40) \\ \hline
\end{tabular}
}
\end{center}
\end{table}

\subsection{Performance evaluation}
In order to validate that the visual imagery-based BCI can be used to control the device, we validated the visual imagery data using CNN with the 10 channels we selected. Table I shows the classification performances of CNN with significant channels. Despite using a single channel, it showed encouraging results between 0.591 and 0.611 in classifying four classes. The highest classification performance was recorded in channel AF3 with 0.611 because blinking and movement are less noticeable. Table II shows the results of classification performance using the combination of two channels with the highest classification performance. As a result, the average classification performance for each subject was 0.610-0.667, which was higher than when one channel was used. As these results are suitable for controlling the robot arm in real life, we conducted an online experiment based on them.

Table III shows the success rate of the online experiment for the three subjects who performed best in the offline experiment. The subjects with the best performance were Sub06, Sub07, and Sub08, and three channel combinations and three runs were performed for each channel combination. Each run included 40 trials, and the lowest success rate was 0.55 and the highest success rate was 0.78. Despite the 0.23 deviation between the highest and lowest success rates, we could investigate the possibility of controlling a visual imagery-based robotic arm according to these results. The combination of AF3--Oz showed the highest classification performance out of the three channel combinations, and its average classification performance was 0.664. This result showed the highest classification performance due to the combination of channels in the frontal and occipital areas. Also, since AF3 performed better than AFz in the frontal area, AF3--Oz had the highest classification performance even when combined with Oz in the occipital area.

\section{CONCLUSION}
In this study, we tested the feasibility of controlling a BCI-based robotic arm in a real environment. Furthermore, we performed a statistical analysis based on neurophysiological data to select significant channels in visual imagery. Based on these results, we evaluated the classification performance using one channel and combinations of two channels. The classification performance with two channels was higher than the classification performance with one channel, and the combination of two channels involving both the frontal and occipital areas had the highest classification performance. Although the results of the online experiment had large deviations for each experiment, we could investigate the feasibility of BCI-based robotic arm control in the real environment.

In future work, we will propose a deep-learning architecture to decode visual imagery with stable performances for BCI-based devices, such as robotic arms that respond to intuitive user intentions. 
Additionally, because EEG data is especially difficult to acquire, augmentation methods will be developed to solve this problem. Finally, we will develop a new conceptual paradigm that increases the degree of freedom of visual imagery to solve the limited command.

\bibliographystyle{IEEEtran}
\bibliography{REFERENCE}

\end{document}